\documentclass[preprint]{aastex}

\shorttitle{Energy dissipation in null points}
\shortauthors{Olshevsky et al.}

\begin{document}


\title{Energy dissipation in magnetic null points at kinetic scales}


\author{Vyacheslav Olshevsky\altaffilmark{1}}
\affil{Centre for mathematical Plasma Astrophysics (CmPA), KU Leuven, Belgium}
\email{sya@mao.kiev.ua}

\and

\author{Andrey Divin}
\affil{Department of Physics, St. Petersburg State University, Russia}

\and

\author{Elin Eriksson\altaffilmark{2}}
\affil{Swedish Institute of Space Physics, Uppsala Division, Uppsala, Sweden}

\and

\author{Stefano Markidis}
\affil{High Performance Computing and Visualization (HPCViz), KTH Royal Institute of Technology, Stockholm, Sweden}

\and

\author{Giovanni Lapenta}
\affil{Centre for mathematical Plasma Astrophysics (CmPA), KU Leuven, Belgium}

\altaffiltext{1}{Main Astronomical Observatory of NAS, Kyiv, Ukraine}
\altaffiltext{2}{Uppsala University, Department of Physics and Astronomy, Uppsala, Sweden}

\begin{abstract}
We use kinetic particle-in-cell and magnetohydrodynamic simulations supported by an observational dataset to investigate magnetic reconnection in clusters of null points in space plasma.
The magnetic configuration under investigation is driven by fast adiabatic flux rope compression that dissipates almost half of the initial magnetic field energy.
In this phase powerful currents are excited producing secondary instabilities, and the system is brought into a state of `intermittent turbulence' within a few ion gyro-periods.
Reconnection events are distributed all over the simulation domain and energy dissipation is rather volume-filling.
Numerous spiral null points interconnected via their spines form null lines embedded into magnetic flux ropes; null point pairs demonstrate the signatures of torsional spine reconnection.
However, energy dissipation mainly happens in the shear layers formed by adjacent flux ropes with oppositely directed currents.
In these regions radial null pairs are spontaneously emerging and vanishing, associated with electron streams and small-scale current sheets.
The number of spiral nulls in the simulation outweighs the number of radial nulls by a factor of 5\---10, in accordance with Cluster observations in the Earth's magnetosheath.
Twisted magnetic fields with embedded spiral null points might indicate the regions of major energy dissipation for future space missions such as Magnetospheric Multiscale Mission (MMS).
\end{abstract}


\section{Introduction}

Magnetic reconnection is a fundamental process believed to be the main mechanism of fast energy release in magnetized astrophysical plasma \citep{Priest:Forbes:2000}.
It is ignited at kinetic scales where ideal conditions for plasma are broken.
At the same scales the energy cascade produced by turbulence becomes dissipative, hence the two processes: turbulence and magnetic reconnection should be mutually connected.

Many scenarios of magnetic reconnection are attributed to the regions where magnetic field vanishes, the null points.
Those are found on the Sun \citep{Longcope:2005LRSP} where they are believed to be the sources of energy release of solar flares \citep{Sweet:1958NC}.
In the Earth's magnetosphere null points are accompanied by various instabilities and wave activity patterns \citep{Xiao:etal:2006NatPh,Wendel:Adrian:2013JGRA,Deng:etal:2009JGRA}.
Detection of null points requires knowledge of the vector magnetic fields implying either polarimetric observations or multi-spacecraft measurements.
Such measurements are not available in the major part of the solar wind yet.
Instead, the tubes of twisted magnetic fields, or magnetic flux ropes are found in solar wind \citep{Moldwin:etal:2000GeoRL,Janvier:etal:2014JGRA}.
The connection of null points and flux ropes, their relation to magnetic reconnection has been unclear, which was a main motive of the present paper. 

Magnetic reconnection in null points has been intensively studied in the framework of magnetohydrodynamics (MHD).
Classification of the null points is based on the linearization of magnetic field: the eigenvalues of the magnetic field gradient define if a null is {\it radial} or {\it spiral} in three dimensions.
Radial and spiral nulls degenerate into X and O points in two dimensions, correspondingly \citep{Lau:Finn:1990,Parnell:etal:1996PhPl}.
MHD theory of magnetic reconnection in null points has been derived by \citet{Priest:Titov:1996RSPSA}, and reconnection regimes were classified by \citet{Priest:Pontin:2009PhPl}.
Three-dimensional modeling of magnetic reconnection in null points has been performed with MHD \citep{Galsgaard:Pontin:2011b} and kinetic \citep{Baumann:Nordlund:2012ApJ} codes.
Most works investigate isolated null points or null pairs, however {\it in situ} observations suggest that null points tend to concentrate in clusters \citep{Deng:etal:2009JGRA,Wendel:Adrian:2013JGRA}.
A study of multiple null points in 3D was presented by \citet{Galsgaard:Nordlund:1997JGR}, however the addressed magnetic configuration was force-free and required artificial external driver to force reconnection, while we are interested in spontaneous reconnection events.
In the recent MHD analysis of \citet{Wyper:Pontin:2014a,Wyper:Pontin:2014b} an initial configuration with a single null point was driven to become unstable.
Plasma then evolved to a turbulent-like state with null point clusters and magnetic flux ropes governing the energy dissipation.

Recently we have proposed a non-conventional non-equilibrium magnetic field configuration where spontaneous magnetic reconnection in null points and flux ropes caused extremely efficient energy dissipation \citep{Olshevsky:etal:2013PhRvL}.
In the subsequent paper \citep{Olshevsky:etal:2015JPP} we have found that plasma was confined in the flux ropes and formed structures similar to Z-pinches.
Efficient energy dissipation in the system was attributed to the secondary magnetic reconnection driven by the instabilities in these pinches.
Here we zoom in on the null points in the same configuration, analyze their topology and associated energy dissipation.
By comparison of kinetic and MHD simulations we have found that the initial relaxation of the system was driven by the large-scale fluid modes.
These modes released a substantial amount of magnetic energy, but brought plasma in the kinetic simulation into a turbulent state within only a few ion gyro-periods.
Investigation of different energy dissipation indicators and power spectra allowed drawing out a possible mechanism of efficient magnetic energy dissipation in space plasmas.
Our idea is that bending and interaction of the flux ropes (with embedded currents and spiral nulls) drive secondary instabilities, for instance at the shear layers between the adjacent ropes.
In such configuration magnetic energy cascades from the large (flux rope) to the smallest (electron) scales where electrostatic turbulent fluctuations are excited that convert magnetic energy to particle heating.

\section{Simulations}

In this section the results of two simulations are discussed: the main subject of this paper, a kinetic PIC simulation described in Appendix~\ref{app:PIC}, and a resistive MHD simulation (Appendix~\ref{app:MHD}).
Two simulations were carried out on the Cartesian grid with the same number of cells, $400^3$, and with the same initial magnetic/gas energy ratio.
The magnetic configuration under study is fully periodic, magnetic field is given by
\begin{eqnarray*}
\nonumber
  B_x & = & -B_0\cos{\frac{2\pi x}{L_x}}\sin{\frac{2\pi y}{L_y}}, \qquad\qquad \\
  B_y & = & B_0\cos{\frac{2\pi y}{L_y}}\left( \sin{\frac{2\pi x}{L_x}} - 
        2 \sin{\frac{2\pi z}{L_z}} \right), \\
  B_z & = & 2 B_0\sin{\frac{2\pi y}{L_y}}\cos{\frac{2\pi z}{L_z}}. \qquad\qquad \\
\label{eq:initial}
\end{eqnarray*}
The simulation boxes had dimensions $L_x=L_y=L_z=20$ in the corresponding code units (ion inertial length $d_i$ in the PIC simulation, Appendix~\ref{app:PIC}).
Although the adimensional units in the MHD code may be scaled to any physical range, it is inappropriate to scale them to $d_i$, because the essence of the MHD model is its large (comparing to, e.g., ion) scales (Appendix~\ref{app:MHD}).
Therefore we don't directly compare times and distances between the two runs.
However, some relative quantities such as the kinetic/magnetic energy ratio can be quantitatively compared.
We have set the initial magnetic field amplitude to $B_0=1$ in the MHD simulation, and to $B_0=0.0127$ in the PIC simulation to match the initial kinetic/magnetic energy ratio.
Initial uniform density was set to unity in each code's units.

\begin{figure}
\begin{centering}
\includegraphics[width=0.7\textwidth]{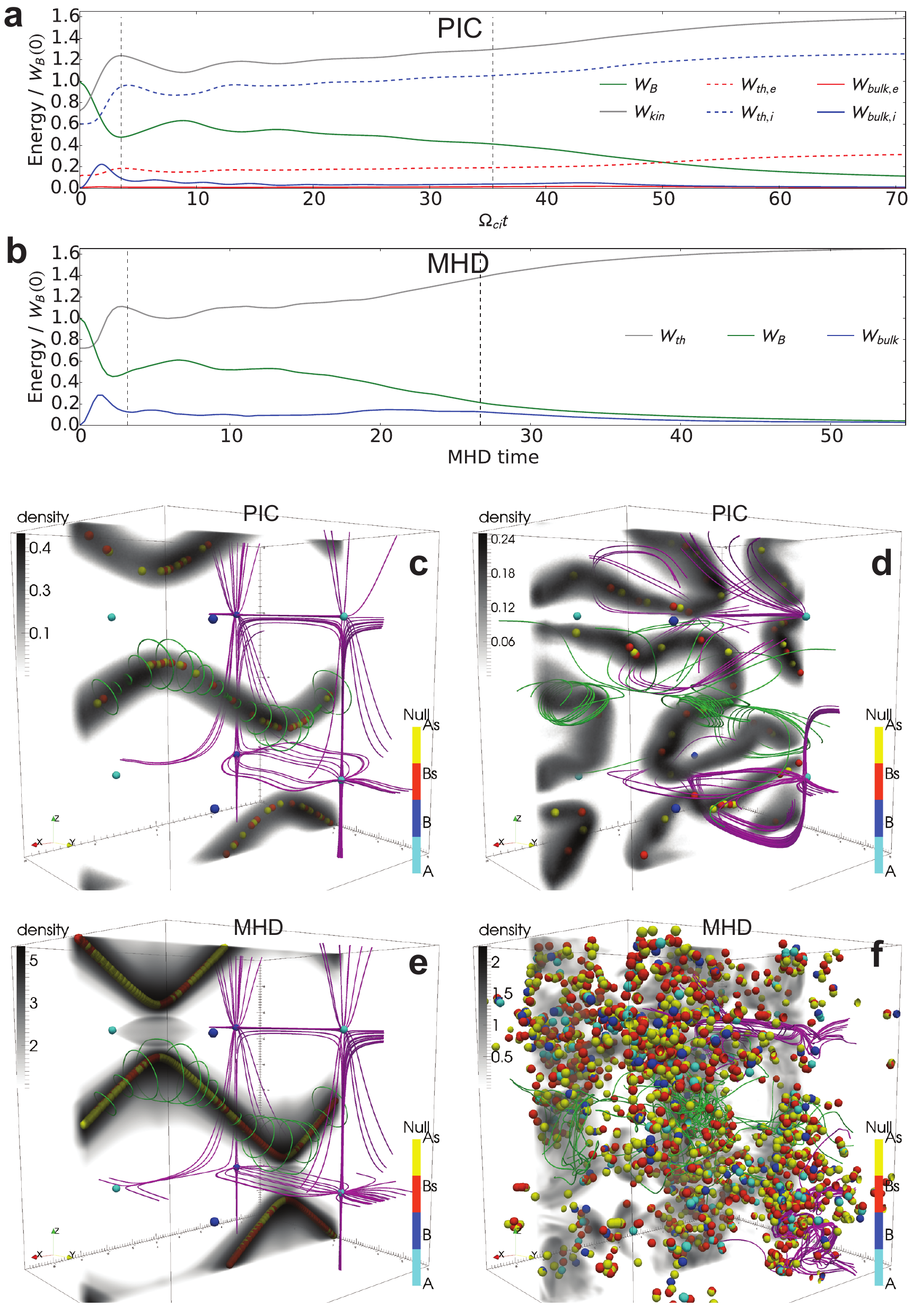}
\caption{Comparison of PIC and MHD simulations.
a, b) Evolution of different components of energy in PIC (a) and MHD (b) runs. Domain-integrated energy components are marked: $W_B$ -- magnetic field; $W_{th}$, $W_{th,e}$, $W_{th,i}$ -- thermal energy in the MHD run, and thermal energies of the two species in the PIC run; $W_{bulk}$, $W_{bulk,e}$, $W_{bulk,i}$ -- same for the bulk kinetic energy. Vertical dashed lines mark the moments at which the snapshots are made.
c, d) Snapshots from PIC simulation corresponding to the two moments marked in (a). Null-points are marked with spheres which colors designate the null point types. Green magnetic field lines are associated with a null line; magenta field lines show the topology of radial nulls. Mass density is shown with shades of grey.
e, f) Snapshots from MHD simulation corresponding to the two moments marked in (b). Meaning of elements is the same as in (c, d).
Note, in (c, d, e, f) the regions adgacent to Y boundaries are masked for clarity, due to the symmetry of the configuration they don't carry additional information.
\label{fig:energy}}
\end{centering}
\end{figure}

The magnetic field configuration described by Equations~\ref{eq:initial} is divergence-free, but not force-free.
Initial density distribution is uniform, hence forces at $t=0$ are not balanced.
Pressure imbalance triggers explosive relaxation during which about a half of the initial magnetic energy is released (Figure~\ref{fig:energy} a, b).
In addition to energy, magnetic field and density distributions are also comparable during this phase as seen from comparison of panels c and e of Figure~\ref{fig:energy}.
Both observations suggest that the initial evolution is driven by large-scale MHD modes.

We use the Poincar\'e index method, as described in Appendix~\ref{app:poincare}, to detect and classify magnetic null points in the simulations.
The classification that we use \citep{Cowley:1973,Lau:Finn:1990} is based on the $\mathbf{\nabla}\mathbf{B}$ eigenvalues, which should sum up to zero due to the divergence-free condition.
When all three eigenvalues are real the null is called {\it radial}.
Radial nulls degenerate into X points in 2D.
When two eigenvalues are complex, the null is called {\it spiral}.
The topology of magnetic field in the vicinity of a spiral null resembles a plasmoid or a helical flux rope; a spiral null degenerates into an O point in 2D.
Further division of nulls is defined by the signatures of the real parts of the $\mathbf{\nabla}\mathbf{B}$ eigenvalues: when two eigenvalues are positive, the null is of B type, also called {\it positive} \citep{Longcope:2005LRSP}; when two eigenvalues are negative, the null is of A type ({\it negative}); similarly, spiral nulls are divided into As and Bs types.
Our initial magnetic configuration contains lines composed of essentially two-dimensional, non-generic O points that are structurally unstable in 3D \citep{Greene:1988}.
The O points evolve into As and Bs nulls just after the beginning of the simulation, and the 3D null lines consist of interconnected spiral nulls As and Bs (Fig.~\ref{fig:energy} c, e).

The energetics of the two simulations is illustrated in panels (a, b) of Figure~\ref{fig:energy}.
The time scale in the MHD run is arbitrary, and the reader should not be confused by the similarity of energy plots: only the initial phases are similar.
Panels (c\---f) of Figure~\ref{fig:energy} display the snapshots of the PIC (c, d) and MHD (e, f) simulations taken at two moments: during the explosive relaxation and at a later stage (marked by the vertical lines in panels (a, b).
In the beginning the null lines are fenced in the helical magnetic field lines.
Because the initial density is uniform, magnetic tension squeezes plasma towards the null lines where high-density regions are formed, as illustrated by the grey shade in panels (c, e) of Figure~\ref{fig:energy} (only 3 out of 9 null lines are shown for clarity).
The process of compression is followed by the reverse expansion, resulting in pulsations visible in kinetic and magnetic energy plots in panels (a, b) during some one-third of the total duration of the simulations.
Notably, the major part of magnetic energy released during the first compression in the PIC simulation is transferred to ions.
These ions create powerful currents along the null lines, and their collective behavior is well approximated by the fluid model.

\begin{figure}
\begin{centering}
\includegraphics[width=0.8\textwidth]{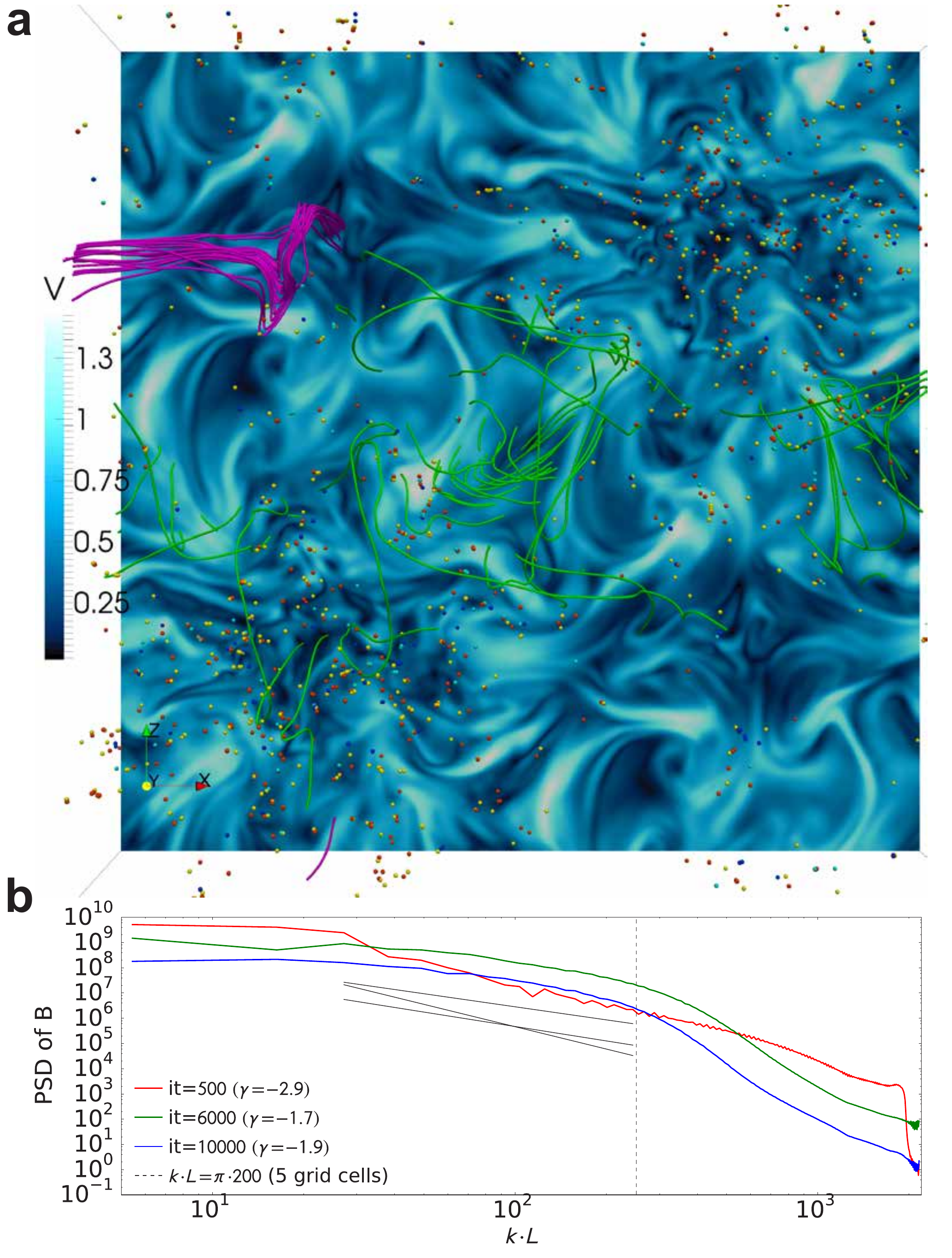}
\caption{Multi-scale chaotic flows in the MHD simulation.
a) Same time moment as in Figure~\ref{fig:energy}~f shown under a different angle. Same magnetic topology illustration, but the null points made smaller for clarity. The $Y=10$ plane is colored with velocity amplitude.
b) Power spectral density (PSD) of magnetic field at three time moments in the MHD simulation.
The color of the spectra indicates the time step index $i=500$ (red), $i=5500$ (green), and $i=8500$ (blue).
The vertical line indicates the wavenumber corresponding to the wavelength of 5 grid cells.
The best-fit exponents are quite far from the expected values in MHD turbulence; the dissipation is purely numerical.
\label{fig:MHD}}
\end{centering}
\end{figure}

When the pressure imbalance is compensated, the behavior of plasma diversifies in the MHD and PIC simulations.
Such dissimilarity is caused by the intrinsic discrepancy between the spatio-temporal scales considered by the two physical models.
In the kinetic simulation there is no numerical resistivity, and the currents along the null lines do not dissipate for relatively long period. Hence the plasma confinement holds, and its behavior is dominated by the current channels \citep{Olshevsky:etal:2015JPP}.
In the MHD simulation the currents are ruined by the resistive dissipation and become disrupted quickly.
In the absence of a large-scale driver, the simulation becomes very chaotic.
Figure~\ref{fig:MHD}a illustrates the flows in the simulation domain at the same moment as in Fig.~\ref{fig:energy}f.
They indeed show no large-scale coherent structures, with vortices of different sizes formed everywhere.
When the pressure imbalance is compensated in the MHD run, the spectrum of magnetic fluctuations (at well resolved scales) gets a power-law shape with exponent $\approx-1.7$ (Fig.~\ref{fig:MHD}b), which corresponds quite well to the measured fluid-scale spectrum in the solar wind \citep{Alexandrova:etal:2012ApJ,Safrankova:etal:2013}.
The fluctuations at scales of a few grid cells are damped by the numerical scheme (slope limiter, see Appendix~\ref{app:MHD}), and the spectrum becomes much steeper.

The situation is quite different in the PIC simulation, where specific scales are defined by ions and electrons.
The absence of numerical resistivity allows currents to hold for relatively long periods, and the confinement holds for tens of ion gyro-periods. 
As shown by \citet{Olshevsky:etal:2015JPP}, Z-pinches are formed along the sinusoidal null lines along which the currents are streaming through twisted magnetic fields.
These pinches play major role in the energetics of the PIC simulation.
Their configuration resembles plasmoids extended to three dimensions \citep{Markidis:etal:2012,Vapirev:etal:2013} and flux ropes observed in the solar wind (see Fig.~\ref{fig:energy}c and Fig.~\ref{fig:spiral}).
Adiabatic compression of the flux ropes driven by the pressure imbalance releases almost half of the initial magnetic energy, but brings the system to a state with a well-defined energy cascade (Section~\ref{sec:PSD}) and numerous small-scale reconnection events governing energy dissipation (Section~\ref{sec:spiral}), within only a few ion gyro-periods.
Null points are important actors in these processes, and it is appropriate to first zoom in, and understand what is happening in nulls during reconnection.
Since magnetic reconnection is a kinetic process, we will only consider the PIC simulation in the remainder of the paper.

\section{Observations}
Of major importance is to understand the relative roles of spiral and radial null points in the processes of energy dissipation in space and solar plasmas.
Certain indications that spiral null points might dominate over the radial ones were found in observations \citep{Xiao:etal:2006NatPh,Wendel:Adrian:2013JGRA,Deng:etal:2009JGRA} and simulations \citep{Wyper:Pontin:2014a}.
Unfortunately, no systematic observational survey has been performed yet: main limitation of the widely employed Poincar\'e index method of null detection (Appendix~\ref{app:poincare}) is its inability to detect nulls outside the spacecraft tetrahedron.
The more advanced method proposed by \citet{Fu:etal:2015JGR} will provide means to perform such a survey \citep{Eriksson:etal:2015}, which we briefly demonstrate in this section.

To fill in the gap and put our study into context, we have applied the two methods, Poincar\'e index and Taylor expansion, to a set of full resolution ($\sim 66$ Hz) DC magnetic field data from the Flux Gate Magnetometer (FGM) \citep{Balogh:2001:FGM} aboard Cluster spacecraft \citep{Escoubet:etal:2001:CLUSTER}.
The data was taken on 27 March 2002 from 09:45 to 11 UT, when the four Cluster spacecraft were in the magnetosheath, downstream of a quasi-parallel bow shock.
The same period was studied before analysed, in particular, by \citet{Retino:etal:2007,Wendel:Adrian:2013JGRA}. 

\begin{figure}
\begin{centering}
\includegraphics[width=0.98\textwidth]{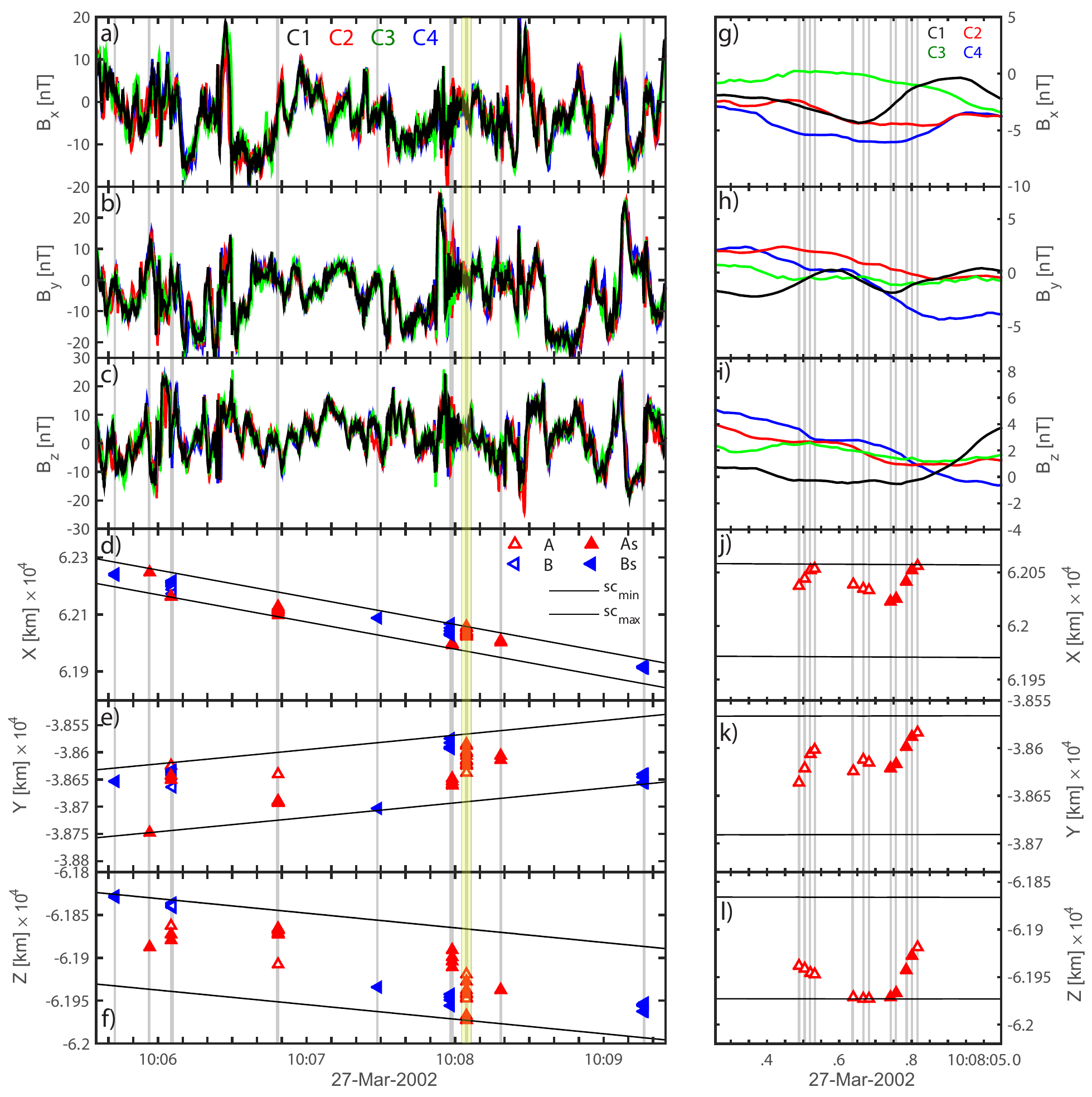}
\caption{Detection of null point in Cluster spacecraft data. 
(a--c) The X, Y, Z components of magnetic field measured by FGM. 
(d--f) The X, Y, Z coordinates of the null points found within error margins (black lines), as computed from equation 1. The $sc_{min}$ and $sc_{max}$ label the minimum and maximum coordinates of the spacecraft along the corresponding axis (mark the edges of the rectangular box of the spacecraft as explained in Appendix~\ref{app:poincare}).
(g--l) Same as (a--f), but for 0.5 second interval around 10:08 UT marked by yellow bar in (a--f).
\label{fig:observations}}
\end{centering}
\end{figure}

Figure~\ref{fig:observations} shows an example of null point detection performed over a 3 minute set of magnetic field data (Panels a--c) on 27 March 2002, around 10:06 UT --- 10:09 UT. 
Positions of the nulls are plotted in panels d--f (X, Y, Z in GSM coordinates, respectively). 
Null points are located with a good accuracy in weak $|B|$ regions, as indicated by error constraint conditions marked by black lines (Appendix~\ref{app:poincare}). 
Notably, positive and negative nulls are commonly found in pairs as visible in Figure~\ref{fig:observations}, 10:06:05 UT, 10:07:58 UT, which is also common in our PIC simulations (see below).

A zoomed view around 10:08:05 UT (Figure~\ref{fig:observations}, g--l) provides an example of transformation of a radial A null into a spiral As null and back. 
At the present stage it is difficult to interpret this result as a clear observation of a single null point conversion. 
The detection procedure assumes linearity of magnetic in the vicinity of a null point, whereas magnetic reconnection simulations reveal the multiscale picture of reconnection with electron beams, plasmoids, thin current layers \citep{Karimabadi:etal:2014}, which can significantly affect remote detection of null points in satellite data.

The numbers of null points of different types are summarized in Table~\ref{tab:nulls}.
The Taylor expansion method has found 443 time steps with nulls, while the Poincar\'e index has found only 64, illustrating the aforementioned limitation of the second method. 
Both methods have detected the dominating (80--90\%) number of spiral nulls, As and Bs. 
There was a slight tendency to have more negative (A and As) nulls than positive (B and Bs) nulls. 

\begin{figure}
\includegraphics[width=0.98\textwidth]{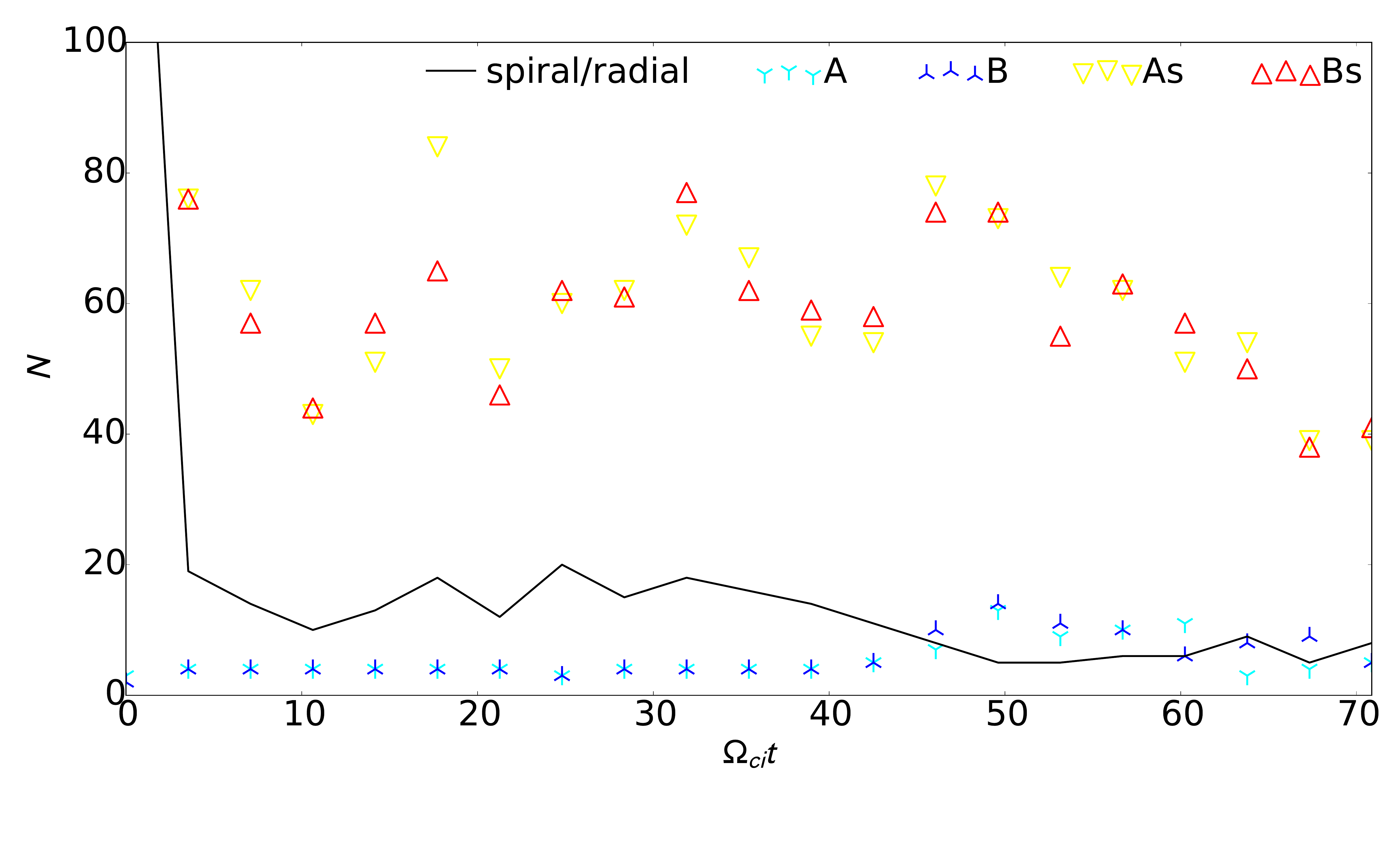}
\caption{Evolution of the number of null points of different types in the PIC simulation. The colors correspond to the null point color legend in Fig.~\ref{fig:energy}.
Two-dimensional O-type nulls exist only in the initial state, often misinterpreted as spiral nulls, the reason for the very high spiral/radial ratio at $t=0$.
\label{fig:null-types}}
\end{figure}

We used Poincar\'e index method as described in Appendix~\ref{app:poincare} to find null points in the simulation domain, the results are shown in Figure~\ref{fig:null-types}.
Our initial magnetic field configuration contains 8 null points of radial A and B types and (mathematically) infinite number of degenerate O-type nulls.
The O-points, initially forming the null lines, evolve into spiral As and Bs nulls as the simulation begins, which explains the sharp drop of the ratio of spiral/radial null numbers.
In such a fashion we artificially create a relaxing system where the spiral nulls outweigh the radial nulls, and the spiral/radial null count ratio keeps high (5\---10) throughout the simulation, in accordance with observations.
\citet{Wyper:Pontin:2014b} concluded the importance of the spiral null points and magnetic flux ropes for magnetic reconnection in their 3D MHD simulations initiated in a completely different magnetic field configuration with a single isolated radial null; clusters of secondary null points were forming during the evolution of such a system.
Therefore, in addition to the observational survey, a numerical investigation should be performed in different field configurations to either support or drop this conclusion.

\section{Spiral nulls}
\label{sec:spiral}
\begin{figure}
\includegraphics[width=0.98\textwidth]{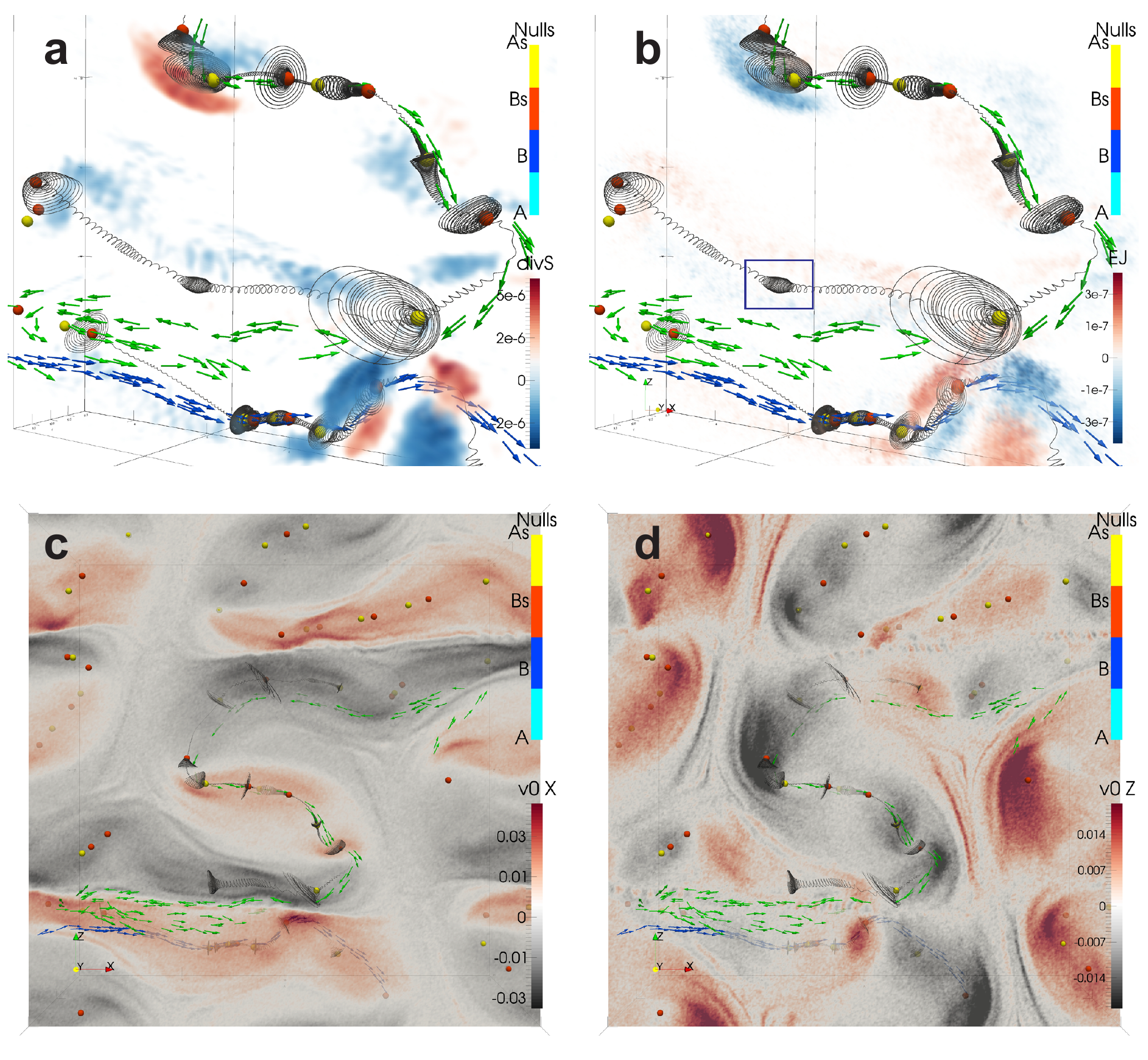}
\caption{Magnetic topology of the spiral nulls at $\Omega_{ci}t=35$.
a, b) A zoom-in on two adjacent flux ropes. Spiraling grey magnetic field reveal the fans of the As and Bs nulls connected via their spines.
Notable is a `knot' in the spiral highlighted by a blue box in (b) that does not contain null points. It may be associated with birth or decay of a null pair.
Green and blue arrows mark electron streamlines in the two oppositely directed current channels. 
Electrons travel along null lines for 10\---20 $d_i$, and then diverge.
The red and blue shade is volume rendering of the divergence of Poynting vector (a) and the magnitude of $\mathbf{E}\cdot\mathbf{J}$ (b).
The most intensive energy conversion happens in the region where the two flux ropes interact or bend.
c, d) A slice through the simulation domain at $Y=10\,d_i$ shows the in-plane electron velocity components $\upsilon_{ex}$ (c) and $\upsilon_{ez}$ (d) with a grey-red pallette.
Null points, velocity vectors and field lines are the same as in the upper panels.
In the lower left corner a shear layer is formed between two current channels.
On the interface a wave-like pattern is visible in (d) that is also seen in the upper-right corner where there is a symmetrical shear layer.
\label{fig:spiral}}
\end{figure}

Previous works have indicated that energy conversion in the discussed configuration is associated with spiral null points connected into null lines embedded into twisted magnetic flux ropes \citep{Olshevsky:etal:2013PhRvL,Olshevsky:etal:2015JPP}.
Figure~\ref{fig:spiral} shows magnetic topology and electron streamlines associated with two adjacent flux ropes with null lines. 
Positive and negative (Bs and As) nulls alternate along the null line. 
The black magnetic field lines in Figure~\ref{fig:spiral} belong to the fan surfaces of the corresponding nulls.
The separators of the adjacent nulls are created by intersecting fans; helical field lines also connect the nulls with the external magnetic field.
This picture is very similar to the one used by \citet{Wyper:Pontin:2014b} to describe the `secondary bifurcations' that lead to the creation of spiral null pairs in their MHD simulations.
Similar topology was observed in a pair of nulls detected by Cluster in the Earth's magnetosheath \citep{Wendel:Adrian:2013JGRA}. 
Reconstruction of magnetic topology of another spiral null pair by \citet{Deng:etal:2009JGRA} revealed a large angle between the spines of the two nulls, and a fan-fan separator line.

Although we haven't performed a dedicated topological analysis of our simulations, the helical field lines surrounding the nulls in Figure~\ref{fig:spiral} a,b suggest their spines to be tangential to the current wires.
MHD theory would classify magnetic reconnection in such configuration as torsional spine reconnection: the currents accumulate along the spines and are co-aligned with them \citep{Priest:Titov:1996RSPSA,Priest:Pontin:2009PhPl}.
In panels (a, b) of Figure~\ref{fig:spiral} a zoom-in on the magnetic topology of the null lines reveals a `knot' of field lines in the central flux rope (highlighted by a blue box in panel b).
No null points are detected in this region, which might indicate an emerging or decaying As-Bs null pair. 
Interestingly, not much energy conversion is associated with this region or the majority of null points directly.
The divergence of the Poynting vector $\mathbf{\nabla}\cdot\mathbf{S}$ and the work of the field $\mathbf{E}\cdot\mathbf{J}$ on particles are high in the regions where adjacent current channels bend or interact with each other, and where clusters of nulls are found.
This observation is supported by other indicators such as electron heating or the violation of frozen-in condition already presented in Figure~4 by \citet{Olshevsky:etal:2015JPP}.
Note, that although the maximum absolute values of $\mathbf{\nabla}\cdot\mathbf{S}$ is substantially higher than the maximum of $\mathbf{E}\cdot\mathbf{J}$, domain-averaged value of the latter quantity is about 10 times larger than the average value of Poynting vector.

The in-plane electron velocities $\upsilon_{ex}$ and $\upsilon_{ez}$ in the slice through the simulation domain show several shear layers where the flux ropes interact or bend (Figure~\ref{fig:spiral}, c, d).
Shear layer indicated by the blue and green electron velocity vectors (same as in panels a, b), and the symmetrical one in the upper right corner of the slice demonstrate oscillatory pattern in $\upsilon_{ez}$.
The wavelength of the oscillation is of the order of $d_i$, however, our preliminary simulations with higher resolution show that these oscillations have fine structure, and they will be addressed in detail in a subsequent paper.

\section{Radial nulls}
\begin{figure}
\includegraphics[width=0.99\textwidth]{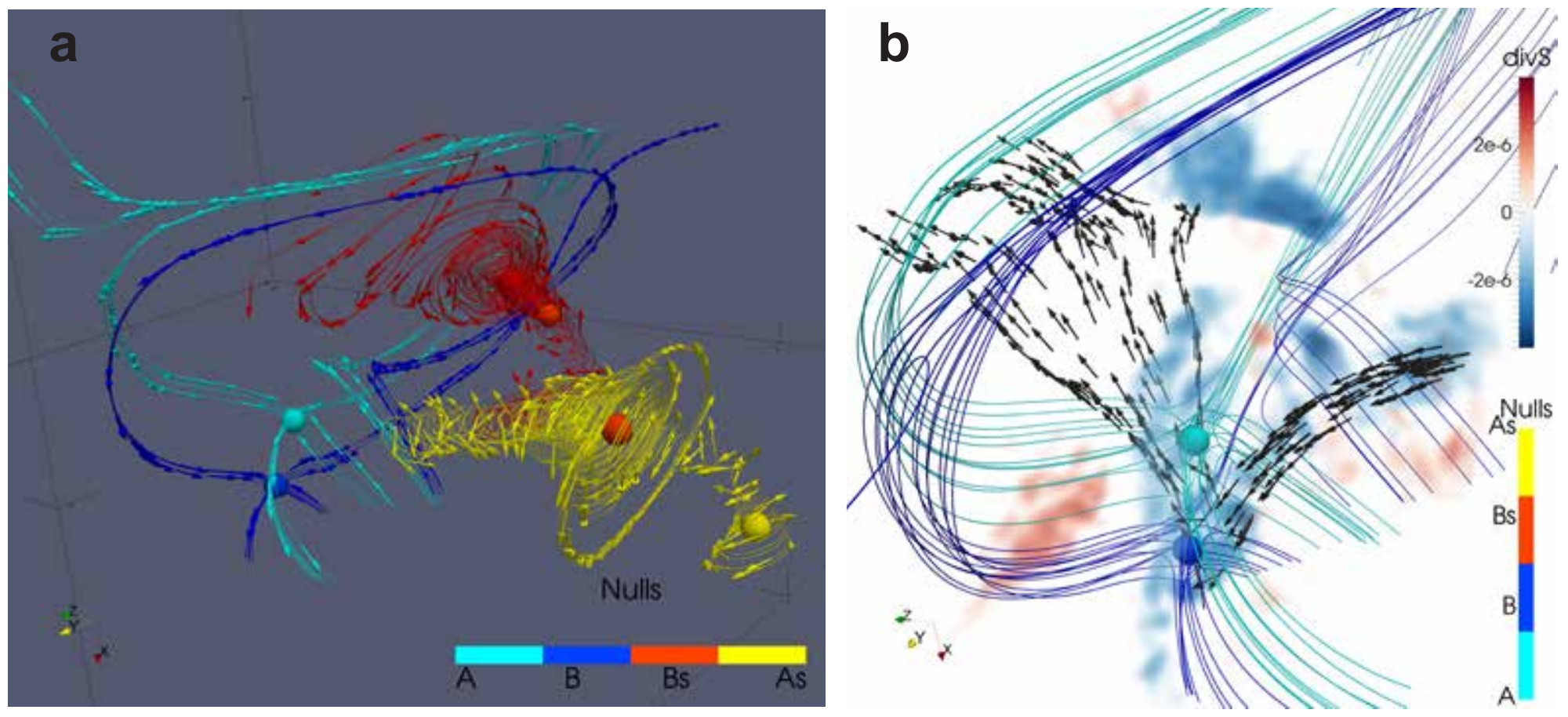}
\caption{Magnetic topology of a radial null pair at $\Omega_{ci}t=42.5$.
a) The A-B null pair (shown with blue and cyan field lines) is formed on the interface between twisted fields surrounding adjacent null lines (marked with red and yellow field lines).
The arrows are magnetic field vectors along the plotted field lines.
b) Same null pair viewed under slightly different angle. Blue and cyan field lines depict the topology of the A and B nulls.
Black arrows mark electron velocity vectors: electron stream enters the B null, bends towards the A null, and scatters away forming a small current sheet surrounded by twisted magnetic field lines.
Red and blue shade marks the regions where the divergence of Poynting vector reaches the highest values. 
As in Figure~\ref{fig:spiral}, these regions correspond to the regions of high $\mathbf{E}\cdot\mathbf{J}$ (with the opposite sign; not shown here).
\label{fig:radial}}
\end{figure}

Figure~\ref{fig:radial} presents a typical A-B null pair spontaneously emerged in the simulation at $\Omega_{ci}t=42.5$ on the interface of the two flux ropes (red and yellow field lines in panel a). 
The distance between the nulls is $1\,d_i$, they disappear in the subsequent simulation snapshot, hence their lifetime is no longer than $7\Omega_{ci}^{-1}$.
The interconnection between these radial nulls is more complex than in the previously shown case of spiral nulls.
The field is non-linear, field lines forming the fans and spines of the nulls are bent and twisted, and reconstruction of such topology from observations would require more sophisticated methods.
Magnetic field lines enter the A null (cyan) in the curved fan surface, and exit it along the spine line, and {\it visa versa} for the B null (blue).
The fan of the A null and the spine of the B null are formed by the same group of field lines that start from the X domain boundary (in the bottom of panel a).
On the bottom, these field lines encircle the yellow flux rope, while upper they twist around the red one, joining the topologies of the two ropes.
The short lifetime of the nulls, their close location, simultaneous appearance and vanishing suggest they are connected by a separator, which, most likely, lies at the intersection of the fan surfaces of the two nulls.

A stream of electrons co-aligned with the spine of the positive (B) null (black arrows in Figure~\ref{fig:radial} b) approaches the null point and reflects towards its negative (A) companion.
After passing the A-point electrons scatter and form a small-scale current sheet surrounded by the twisted magnetic field lines. 
This process is supported by the positive work of the field on the particles and generation of EM fluctuations in the region where the divergence of Poynting vector is negative. 
The described null pair is the first among several A-B pairs created at a later time (see Figure~\ref{fig:null-types}).
Notably, not always null points are born and die in pairs in our simulations and observations: at certain moments the number of positive nulls deviate from the number of negative nulls.
Theory predicts conservation of the topological degree, i.e., each new or disappeared positive null must have its negative counterpart and vice versa.
In practice errors occur on the stage of null detection due to the noise in magnetic field measurements and finite resolution of the numerical grid or spacecraft instruments.
Additional errors are introduced during null classification, especially when the eigenvalues of the $\mathbf{\nabla B}$ have very small imaginary (real) parts.
Influence of these errors on the observations of null points are addressed in \citet{Fu:etal:2015JGR,Eriksson:etal:2015}, see also Appendix~\ref{app:poincare}.

\section{Energy cascade}
\label{sec:PSD}
As we pointed out in previous sections, energy dissipation in our simulation is associated with electron and ion beams streaming through null points, small-scale curent sheets and sheared motions.
Especially important the large-scale flows are in the first phase of ``implosive'' relaxation when the energy of ion beams is about a half of the magnetic field energy (see Fig.~\ref{fig:energy}, a).
Two-stream and shearing instabilities associated with these flows are expected to drive electrostatic turbulent fluctuations and lead to creation of energy cascade.
A somewhat better resolved 3D kinetic simulation \citep{Daughton:etal:2011NatPh} of flux ropes produced by magnetic reconnection in a Harris sheet \citep{Harris:1962NC} has indeed shown `intermittent' turbulence, analyzed in detail by \citet{Leonardis:etal:2013PhRvL}.
Of course, our simulations do not have external driver which is typically present in turbulent studies, and, at our best, we can attribute it as a case of intermitent turbulence.

\begin{figure}
\includegraphics[width=0.99\textwidth]{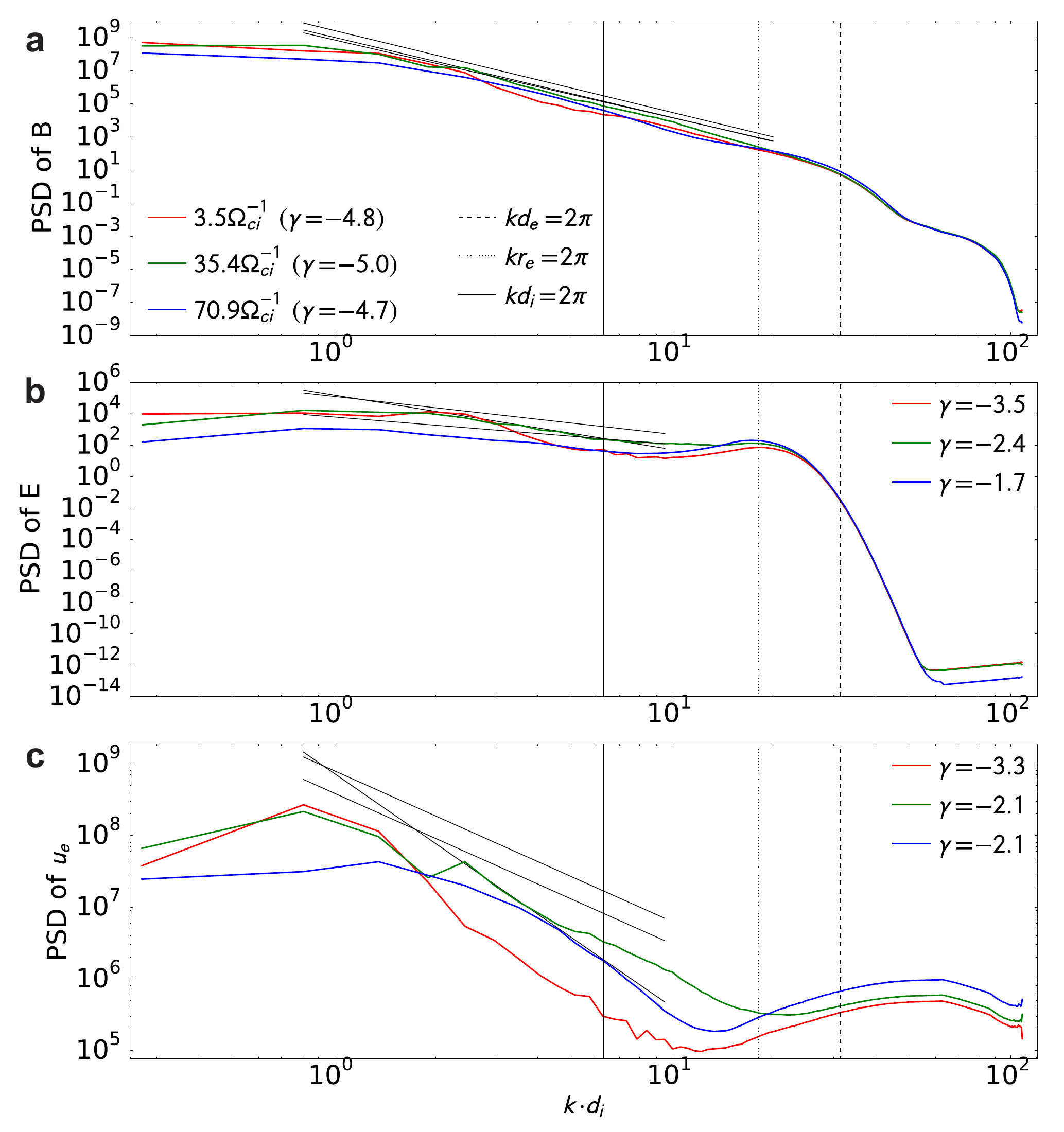}
\caption{Spectra of magnetic field (a), electric field (b) and electron velocity (c) fluctuations at three moments in the simulation.
The color of the spectra indicates the time moment $\Omega_{ci}t=3.5$ (red), $\Omega_{ci}t=35$ (green), and $\Omega_{ci}t=71$ (blue).
The vertical lines indicate $kd_e=2\pi$ (dashed), $kd_i=2\pi$ (solid), and $kr_e=2\pi$ (dotted) computed at the end of the simulation basing on the electron thermal speed and domain-averaged magnetic field value.
The energy from large-scale magnetic features is transferred into small-scale motions and heat (maximum on the right of the panel c).
To stabilize the numerical solution, electric field is smoothed beyond electron gyroscales (sharp drops in panels a, b around the dashed line).
\label{fig:spectra}}
\end{figure}

Indeed, in a few ion gyro-periods magnetic energy spectrum obtains a characteristic shape (red curve in Fig.~\ref{fig:spectra}, a) with four distinct breaks.
The first break at $kd_i=1$ is a transition from fluid to kinetic (ion) scales, similar to the one observed in solar wind \citep{Alexandrova:etal:2012ApJ}.
Unfortunately our simulation box is too small to properly represent the fluid range in the spectrum.
The second break is associated with electron inertial scales $kd_e=2\pi$ and also corresponds to the one detected in solar wind by \citet{Sahraoui:2013ApJ}.
The magnetic energy spectrum between $kd_i=1$ and $kd_e=1$ is rather smooth power-law with a best-fit exponent $\gamma\approx-5$ throughout the simulation.
Such steep shape is found at the sub-electron scales in solar wind \citep{Sahraoui:2013ApJ}.

The third break in the $B$ spectrum $kd_i\approx 50$ makes it milder, while the fourth steepens it further, in discordance with the measurements of \citet{Sahraoui:2013ApJ}.
These scales, however, correspond to the wavelengths shorter than 2 grid points, where the implicit scheme is damping numerical noise.
Panel b of Figure~\ref{fig:spectra} shows the spectra of electric field fluctuations.
To damp numerical noise and stabilize the solution, smoothing is applied to the electric field derived at each time step.
The smoothing operates at five-point stencil, and small-scale $E$ fluctuations are damped to machine precision.
The corresponding break is seen in the electric field spectra at $kd_i\approx20$.
Although smoothing is applied only to electric field, which energy budget is substantially smaller than that of the magnetic field, it might also affect the shape of the magnetic spectrum at the smallest scales.

The distinct features of the electric energy spectrum at $\Omega_{ci}t=3.5$ are the hump at ion inertial scale, followed by a local minimum at $kd_i=2\pi$, and another hump at $kr_e=2\pi$ (electron gyroradius $r_e$ is computed from the domain-averaged average magnetic field and electron thermal speed at $t\Omega_{ci}=70.9$).
The first hump disappears at later time moments, and we associate it to the initial compression of the magnetic flux ropes, excitation of ion currents and acceleration of ions to non-thermal speeds \citep{Olshevsky:etal:2015JPP}.
The peak at electron scales that may be blurred due to the aforementioned smoothing. 

The spectra of electron velocity fluctuations (Figure~\ref{fig:spectra}, c) have a well defined peak at the smallest, sub-electron scales at all time moments.
The energy contained in these scales increases with time, indicating heating of electrons due to the dissipation of magnetic energy at the smallest scales.
Notably, the power of velocity fluctuations and electric field fluctuations start increasing at frequency $kd_i=10$.
The higher frequency fluctuations of electric field, however, are damped by numerical smoothing.
The variation of the velocity spectra with time allows to conclude that in the beginning (red) most of the energy is contained in the current channels with diameters of about $d_i$; later (green and blue), the energy redistributes from the large scales to the small scales, with a distinct local minimum between ion and electron scales.
In the absence of (externally driven) large-scale fluid motions the energy from ion-scale magnetic structures is redistributed to electron scales, where electron velocity fluctuations are excited.

\section{Summary}

We have investigated magnetic reconnection in a cluster of null points by means of kinetic particle-in-cell simulation.
The simulation is initiated from an unbalanced configuration, abrupt relaxation of which excites powerful currents and brings the system to a turbulent state within a few ion gyro-periods.
Major role during relaxation is played by adiabatic compression of magnetic flux ropes: it excites powerful currents along the embedded null lines, and accelerates ions to suprathermal speeds.
Large-scale fluid oscillations are excited at this stage, as confirmed by an accompanying MHD simulation performed in the same magnetic configuration.

When the pressure imbalance is compensated, confinement is ruined in the MHD run, and the system becomes very turbulent.
Energy spectra obtains a distinctive power-law spectra with exponent $\approx-1.7$ in the well-resolved range, becoming steeper at the shorter wavelengths where the dissipation is caused by numerical scheme.
Multiple null points are created all over the simulation domain, but an MHD simulation doesn't allow for an insight into the micro physics of magnetic reconnection in turbulence.

In the PIC simulation distinct spatial scales are dictated by particles, plasma confinement is preserved, and the dynamics in the stationary dissipation phase is dominated by interacting magnetic flux ropes.
The flux ropes in our simulation are formed along null lines, the sequences of spiral null points connected via their spines.
Magnetic reconnection in these null points can be classified as spine reconnection, with energy being transferred to particles mainly in the regions of flux rope bending or interaction.
On the interfaces of oppositely directed adjacent flux ropes the shear layers are created that display oscillatory pattern with a wavelength of the order of ion inertial length. 
A simulation with higher spatial resolution is required to analyze this instability.

The number of null points in the simulation varies with time, with null pairs spontaneously emerging and vanishing.
A typical pair of radial null points is created on the interface of the two interacting flux ropes. 
Such pair is usually short-lived, it reconnects in a few ion gyro-periods.
The distance between the nulls in the newborn pair is about one ion inertial length.
Visualization of the local magnetic topology suggests that the separator of the pair lies in the fan surfaces of the nulls.
Magnetic reconnection in such pair is characterized by a stream of electrons entering the system along the spine of the positive null, refracting towards the negative null, and then scattering away in a small-scale current sheet.

Magnetic reconnection changes the number of null points in the simulation domain, however there always are 5\---10 times more spiral than radial nulls.
We have applied two methods of null point detection: Poincar\'e index and Taylor expansion, to a one-hour period of Cluster measurements in the Earth's magnetosheath.
The second method has found seven times more nulls (443) than the first one (64), and in both cases more than $80$\% of nulls were spiral.
For a spiral null pair from the same period of Cluster observations \citet{Wendel:Adrian:2013JGRA} have deduced magnetic field topology and revealed a spine-aligned current, which corresponds well to our simulations.

Our kinetic simulations, supported by the results of {\it in situ} observations in the magnetosheath suggest a new mechanism of magnetic energy dissipation in turbulent space plasma.
It is governed by bending and interaction of magnetic flux ropes that produce the cascade of magnetic energy to the sub-electron scale, where small-scale electron fluctuations are excited.
Electron velocity fluctuation spectra gets a characteristic maximum at sub-electron scales.
Locally, the dissipation is attributed with clusters of interconnected spiral null points with co-aligned current ropes, short-living pairs of radial nulls, shear layers and small-scale current sheets.

\acknowledgments

Authors are thankful to Yu. Khotyaintsev, A. Vaivads, and H. Fu for useful discussions.
This research has received funding from the European Commission's FP7 Program with the grant agreement eHeroes (project 284461, www.eheroes.eu).
E.E. and S.M are supported by the Swedish VR grant D621-2013-4309.
The simulations were conducted on the computational resources provided by the PRACE Tier-0 projects 2011050747 (Curie) and 2013091928 (SuperMUC).

\appendix

\section{PIC simulation}
\label{app:PIC}

The kinetic simulation of collisionless plasma was carried out using semi-implicit fully electromagnetic PIC code iPic3D \citep{markidis:etal:2010}.
The code solves the time-dependent Maxwell equations for the fields given on a stationary grid, and the equations of motion for computational particles derived from Vlasov equation.
Each computational particle represents a blob of real particles (ions and electrons) that are close to each other in 6D phase space.
The physical units in the code are normalized to the corresponding plasma parameters: proton inertial length $d_i$, proton plasma frequency $\omega_{pi}$, and proton mass $m_i$, hence magnetic field unit is $m_i\omega_{pi}/e$.

The simulation was carried out in a cubic box of size $20\, d_i$ and the resolution of $400$ cells in each dimension.
Two species, ions and electrons were considered, with mass ratio $m_i/m_e=25$, and 64 particle of each specie per cell.
The time step was set to $0.15\omega_{pi}$, satisfying the finite-grid stability criterion; the total duration of the run was $70$ ion gyro periods $\Omega_{ci}^{-1}$.
Particles were initiated with a Maxwellian distribution with thermal speed in each dimension $u_{th,e}=0.02$ for electrons, and $u_{th,i}=0.0089$ for ions, which corresponds to the temperature ratio $T_i/T_e=5$ typical in the Earth's magnetosheath plasma.

Initial domain-integrated magnetic/thermal energy ratio was set to $W_{mag}/W_{th}=1.38$ and decreased to $0.07$ in the end of the simulation, passing through the stages of different plasma $\beta$.
Given these parameters, the electron inertial scales were resolved substantially well, however to resolve electron gyroradius at all stages and to study the fine structure of induced instabilities, higher resolution simulations or mesh refinement techniques \citep{Innocenti:etal:2015} are required.

\section{MHD simulation}
\label{app:MHD}

A total variation diminishing Lax-Friedrichs (TVDLF by \citet{Toth:Odstrcil:1996JCP}) scheme with explicit time stepping (second-order accurate in space and time) was implemented in the code infrastructure of iPic3D (hereinafter referred to as `iPic3D-MHD') based on the serial MHD code by \citet{Kiehas:etal:2006}. 

iPic3D-MHD solves conventional resistive MHD equations cast in dimensionless form (see, e.g., \citet{Ma:Bhattacharjee:2001}). 
The ideal MHD equations were used for the present study 
\begin{eqnarray*}
\nonumber
\frac{\partial\rho}{\partial t} + \mathbf{\nabla}\cdot\left( \rho\mathbf{\upsilon} \right) = 0, \qquad\qquad\qquad \\
\frac{\partial}{\partial t}\left( \rho\mathbf{\upsilon} \right) + \mathbf{\nabla} \cdot \left( \mathbf{\upsilon}\rho\mathbf{\upsilon} -\mathbf{BB}  \right) + \nabla\left( p + \frac{B^2}{2} \right) = 0, \\
\frac{\partial e}{\partial t} + \mathbf{\nabla}\cdot\left( \mathbf{\upsilon}e + \mathbf{\upsilon}\left( p + \frac{B^2}{2} \right) - \mathbf{BB} \cdot \mathbf{\upsilon} \right) = 0, \\
\frac{\partial \mathbf{B}}{\partial t} + \mathbf{\nabla}\cdot \left( \mathbf{\upsilon B} - \mathbf{B \upsilon} \right) = 0, \qquad\qquad \\
e = \frac{p}{\gamma - 1} + \frac{\rho\upsilon^2}{2} + \frac{B^2}{2},\qquad\qquad
\end{eqnarray*}
where $e$ is the total energy density, $\gamma=5/3$ is the specific heats ratio.
The energy components shown in Figure~\ref{fig:energy}b are given by: magnetic field energy $W_B = \frac{1}{2}\int B^2dV$, thermal energy $W_{th} = \frac{1}{\gamma-1}\int pdV$, bulk flow energy $W_{bulk} = \frac{1}{2}\int \rho \upsilon^2dV$.

The numerical diffusivity required for the stability of the simulation was provided by TVD slope limiters \citep{Toth:Odstrcil:1996JCP}.
We have tested all three slope limiters proposed by \citet{Toth:Odstrcil:1996JCP}, and haven't found a notable influence on the energetics of the simulation.

Magnetic field at time $t=0$ was set according to Equations~\ref{eq:initial} using $B_0=1$ in dimensionless units. 
Plasma density $\rho=1$ and pressure $p$ were set uniform at $t=0$.
Initial plasma pressure $p$ equals to $p_i+p_e$, the sum of the initial ion and electron pressures in the PIC simulation ($(m_e/m_i)u^2_{th,e} + u^2_{th,i})/B_0^2=0.6$).
The integrated initial magnetic to thermal energy ratio was $W_B/W_{th}=1.38$, same as in the PIC simulation.

The divergence-free condition for magnetic field ($\mathbf{\nabla B}=0$) is satisfied at t=0. 
However, the B field can slowly accumulate magnetic charges over the course of the run, unless corrected. 
Numerical magnetic field divergence is cleaned each 10$^{th}$ time step by means of the projection method \citep{Toth:2000}.

A uniform Cartesian grid (400 x 400 x 400 cells) with equal spacings in all three dimensions was used. The number of grid points was identical to that in the PIC simulation (Appendix A) to ease the comparison.

\section{Location and classification of null points}
\label{app:poincare}
The problem of locating a magnetic null is essentially a problem of finding a root of a continuous divergence-free vector field.
The widely adopted method of null point location is based on the topological degree, or Poincar\'e index, and was introduced by \citet{Greene:1992}.
A topological degree estimates a number of roots (nulls) in a closed region of space; it is non-zero when an odd number of roots are enclosed in the region.
Commonly used assumption is that the separation between spacecraft is so small that the non-zero topological degree corresponds to exactly one null.

The method introduced by \citet{Greene:1992} used a rectangular box as a volume for which the topological degree was computed.
However, since Cluster mission consists of four spacecraft, the method was adjusted to find nulls in a tetrahedron.
Null detection in the simulation data was also performed with the same method for consistency. 
Each cubic mesh cell was divided into five tetrahedrons with vertices at the mesh nodes.

Computation of the topological degree is based on finding a solid angle between vectors in 3D.
Unlike other implementations we used the formula for solid angle enclosed by 3 vectors proposed by \citet{VanOosterom:Strackee:1983IEEE}; it appears faster and more stable than the traditional implementation based on the Cosine theorem. 
In particular, there is no need for zero-denominator checks when using a present-day programming environment: errors are handled by the arctan2 function.

The magnetic field is assumed to be linear in the vicinity of a null point.
Linear expansion of the magnetic field is a basis for the classification of nulls introduced by \citet{Cowley:1973,Greene:1988,Lau:Finn:1990}.
The eigenvalues of the $\mathbf{\nabla}\mathbf{B}$ tensor define the topological type of the null.
We detect the null point type by estimating the $\mathbf{\nabla}\mathbf{B}$ as described in \citet{Khurana:etal:1996IEEE}.

Obviously, a shortcoming of the Poincar\'e index method is that it can only detect a null point inside the tetrahedron.
This is fine in the models where the ``spacecraft'' cover the whole volume of the simulation domain, but is a severe limitation in real observations.
Therefore, to the latter, we have also applied a more advanced method based on Taylor expansion of the magnetic field in the vicinity of a null \citep{Fu:etal:2015JGR}:
\begin{equation}
\textbf{B}\left(\textbf{r} \right) = \nabla \textbf{B} \left( \textbf{r} - \textbf{r}_0 \right) ,
\end{equation}
where $\textbf{r}$ is the location in space, and $\textbf{r}_0$ is the location of the null. 
Inversion of this linear expansion gives the null position $\textbf{r}_0$. 
In general, the equation will always give a position of a magnetic null. 
However, we regard the null reliably identified only if it is located in a rectangular box defined by the spacecraft location. 
The edges of the box in each direction (x, y, z) are given by the maximum and minimum positions of all Cluster spacecraft. 
Only the magnetic null positions found within this box are type identified and further evaluated.

We have used two error constraints to estimate the accuracy of null point detection: 
\begin{equation} 
\left| \frac{ \nabla \cdot \textbf{B}}{max( \nabla \textbf{B})}\right|, 
\left| \frac{ \lambda_{1} + \lambda_{2} + \lambda_{3} }{ max(| real(\lambda) | )} \right| \; .
\end{equation}
Possible errors of null detection via Taylor expansion are addressed in \citet{Fu:etal:2015JGR}.
For the purpose of the current work, we have tested four thresholds set for both error conditions (i.e., when at least one error condition is above the threshold, the null is discarded): 10, 20, 30, and 40\%. As expected, we found no notable difference in the ratio of the spiral/radial null numbers: indeed, the threshold only limits the number of the selected data points, not their type.
Table~\ref{tab:nulls} presents the results for the error threshold of 40\%.

\bibliographystyle{apj}
\bibliography{3dnull}

\clearpage

\clearpage

\begin{table}[H]
\centering
 \begin{tabular}{c|c c c c c}
  & Nulls found & A (\%) & B (\%) & As (\%) & Bs (\%)\\ 
  \hline 
  Poincar\'{e} index & 64 & 8 & 1 & 55 & 36\\ 
  Taylor expansion & 443 & 14 & 8 & 42 & 36\\ 
  \end{tabular} 
  \caption{Percentage of magnetic nulls of different types detected with Poincar\'{e} index and Taylor Expansion methods in Cluster measurements in Earth's magnetosheath.}
  \label{tab:nulls}
\end{table}





\end{document}